\documentclass[conference]{IEEEtran}
\IEEEoverridecommandlockouts

\usepackage{cite}
\usepackage{graphicx}
\usepackage{epstopdf}
\usepackage{diagbox}
\usepackage{multirow}
\usepackage[cmex10]{amsmath}
\usepackage[ruled]{algorithm2e} 
\usepackage{url}
\usepackage{lineno,hyperref}
\usepackage{amsthm}
\usepackage{booktabs}
\usepackage{subfigure}
\usepackage{url}
\usepackage{mathrsfs}
\usepackage{amsmath}
\usepackage{amsthm}
\usepackage{bbding}

\usepackage{amsfonts,amssymb}
\usepackage{mathrsfs}
\usepackage{amsmath}
\usepackage{amsthm}
\usepackage{bbding}
\usepackage{cite}
\usepackage{xcolor}
\usepackage{bm}

\usepackage{amsmath,amssymb,amsfonts}
\usepackage{algorithmic}
\usepackage{graphicx}
\usepackage{textcomp}
\usepackage{url}
\usepackage{multirow}
\usepackage{booktabs}
\usepackage{lineno,hyperref}
\usepackage{xcolor}
\def\BibTeX{{\rm B\kern-.05em{\sc i\kern-.025em b}\kern-.08em
    T\kern-.1667em\lower.7ex\hbox{E}\kern-.125emX}}
\begin{document}

\title{Robust Divergence Learning for Missing-Modality Segmentation
\thanks{Identify applicable funding agency here. If none, delete this.}
}
\author{\IEEEauthorblockN{Runze Cheng$^{1,2}$, Zhongao Sun$^{1}$, Ye Zhang$^{1,3}$, and Chun Li$^{1,*}$}
\IEEEauthorblockA{\textit{$^{1}$MSU-BIT-SMBU Joint Research Center of Applied Mathematics, Shenzhen MSU-BIT University, Shenzhen, 518172, China.} \\
\textit{$^{2}$Institute of Control Theory and Control Engineering, School of Automation, Beijing Institute of Technology, 100081, Beijing, China.}\\
\textit{$^{3}$School of Mathematics and Statistics, Beijing Institute of Technology, 100081, Beijing, China.}\\
Email: 3120220916@bit.edu.cn; sunzhongao0224@gmail.com; ye.zhang@smbu.edu.cn. \\
*Corresponding author: Chun Li (E-mail: lichun2020@smbu.edu.cn).}
}

\maketitle

\begin{abstract}
Multimodal Magnetic Resonance Imaging (MRI) provides essential complementary information for analyzing brain tumor subregions. While methods using four common MRI modalities for automatic segmentation have shown success, they often face challenges with missing modalities due to image quality issues, inconsistent protocols, allergic reactions, or cost factors. Thus, developing a segmentation paradigm that handles missing modalities is clinically valuable. A novel single-modality parallel processing network framework based on Hölder divergence and mutual information is introduced. Each modality is independently input into a shared network backbone for parallel processing, preserving unique information. Additionally, a dynamic sharing framework is introduced that adjusts network parameters based on modality availability. A Hölder divergence and mutual information-based loss functions are used for evaluating discrepancies between predictions and labels. Extensive testing on the BraTS 2018 and BraTS 2020 datasets demonstrates that our method outperforms existing techniques in handling missing modalities and validates each component's effectiveness.
\end{abstract}


\section{Introduction}
Brain tumors are aggressive diseases requiring early detection for effective treatment. MRI is widely used for evaluating brain tumors due to its superior soft tissue contrast and lack of radiation exposure. MRI segmentation is crucial for isolating healthy tissue from abnormal cells, offering various modalities for effective tumor detection, including T1-weighted, T1-weighted post-contrast- enhancement, T2-weighted, and FLAIR. Several existing methods \cite{nnunet,multimoseg,sslswin, p55,p56,p57} achieved high accuracy in tumor segmentation when all modalities are available. However, in real-world scenarios, one or more modalities may be missing due to patient movement, hardware issues, or other factors. This “missing modality” problem arises when one or more modalities (e.g., T1w, T2, T1c, and FLAIR) are missing during inference but available during training \cite{surveyOnMiss}.

Several approaches have been developed to address this issue, which can be categorized into two types \cite{M3AE}: modeling each missing case individually or using a single model to handle all cases. For the former, knowledge distillation is commonly used to transfer competencies from a well-trained teacher model to a student model designed for specific missing modalities. SMU-Net \cite{Azad2022SMUNetSM} employed a novel distillation strategy where a multimodal teacher network transfers knowledge to unimodal student networks at both the latent space and network output levels. ProtoKD \cite{Wang2023} integrated prototype learning with knowledge distillation, effectively capturing the underlying data distribution. MMCFormer \cite{karimijafarbigloo2023mmcformer} leveraged transformers with auxiliary tokens to facilitate modality-specific representation transfer.

The latter category aims to address all missing-modal situations with a single model, typically involving separate modality encoders to project each modality into a shared latent space before feature fusion. RFNet \cite{rfnet} integrated characteristics from various sources using a region-cognizant component. Moreover, Ting and Liu \cite{Multimodal} used modality-specific encoders, a shared decoder, and a strategy to complement incomplete data with complete data. Furthermore, Wang et al. \cite{cvpr2023shapedmodel} employed specific and shared encoders to handle missing modalities for both segmentation and classification tasks.

However, these approaches have some shortcomings. Using a specific model for each missing modality scenario is training-costly, for example, $2^N-1$ models need to be trained when there are $N$ modalities\cite{M3AE}. Conversely, a single model for all cases often results in performance deficiencies with few modalities available \cite{Azad2022SMUNetSM} and high inference costs due to numerous parameters.

Inspired by Chang et al. \cite{36} and  high mutual information knowledge transfer learning \cite{p52}, we process four different modalities individually to preserve unique information and enhance the model's ability to recognize diverse data features, which can handle all cases with signal model with shared backbone. Specifically, we propose a novel mutual information-based metric with Hölder divergence\cite{37} that evaluate discrepancies between the predictions and labels. What is more, a dynamic sharing framework is introduced, which allows the model to adapt its parameters depending on the availability of different modalities. 

The main contributions of this paper are summarized as follows: 1. Novel Network Architecture: We propose a new network architecture for parallel computing based on 3D U-Net. This framework combines unimodal parallel processing and dynamic network module combinations to handle missing modalities during brain tumor segmentation training. 2. New Metric Introduction: the Hölder divergence and mutual information are introduced to evaluate discrepancies between model predictions and labels. By minimizing the distances, we achieve more accurate feature alignment. 3. Extensive Validation: Extensive experiments on the BraTS 2018 and BraTS 2020 medical image datasets \cite{p38} are conducted. The results demonstrate that our method achieves state-of-the-art performance, showcasing its efficiency and practicality. The main notations used in this work is shown in Table \ref{tab-001}.

\begin{table}[t]
	\centering
	\caption{Main Notations Used in This Work. This table provides an overview of the main notations used throughout this work, offering a concise reference for understanding the symbols and terminology employed in the algorithms discussed.}
	\setlength{\tabcolsep}{3pt}
	\begin{tabular}{p{1.8cm}<{\raggedright}|p{6cm}<{\raggedright}} 
		\toprule [1.0pt]
		Notation&Definition\\
		\midrule [0.5pt]
		$x_i$&  the $i^{th}$ modality data of the sample. \\
		$d_i$ & The deepest-level feature of the $i^{th}$ modality. \\
		$d_f$ & The deepest-level full-modality feature. \\
		$h_i$ & The generated single-modality representation of the $i^{th}$ 
modality. \\
		$\widehat{Y}$ & Integrated output under missing modalities. \\
        $Y$ & real sample. \\
        $p(d_f\mid d_m)$ & The conditional distribution of the feature f given the missing modality information m.\\
        $q(d_f\mid d_m)$ & The conditional distribution approximated using variational methods.\\

		\bottomrule[1.0pt]
	\end{tabular}
	\label{tab-001}
\end{table}

\begin{figure*}[hbt!]
	\centering
	\includegraphics[width=0.8\linewidth]{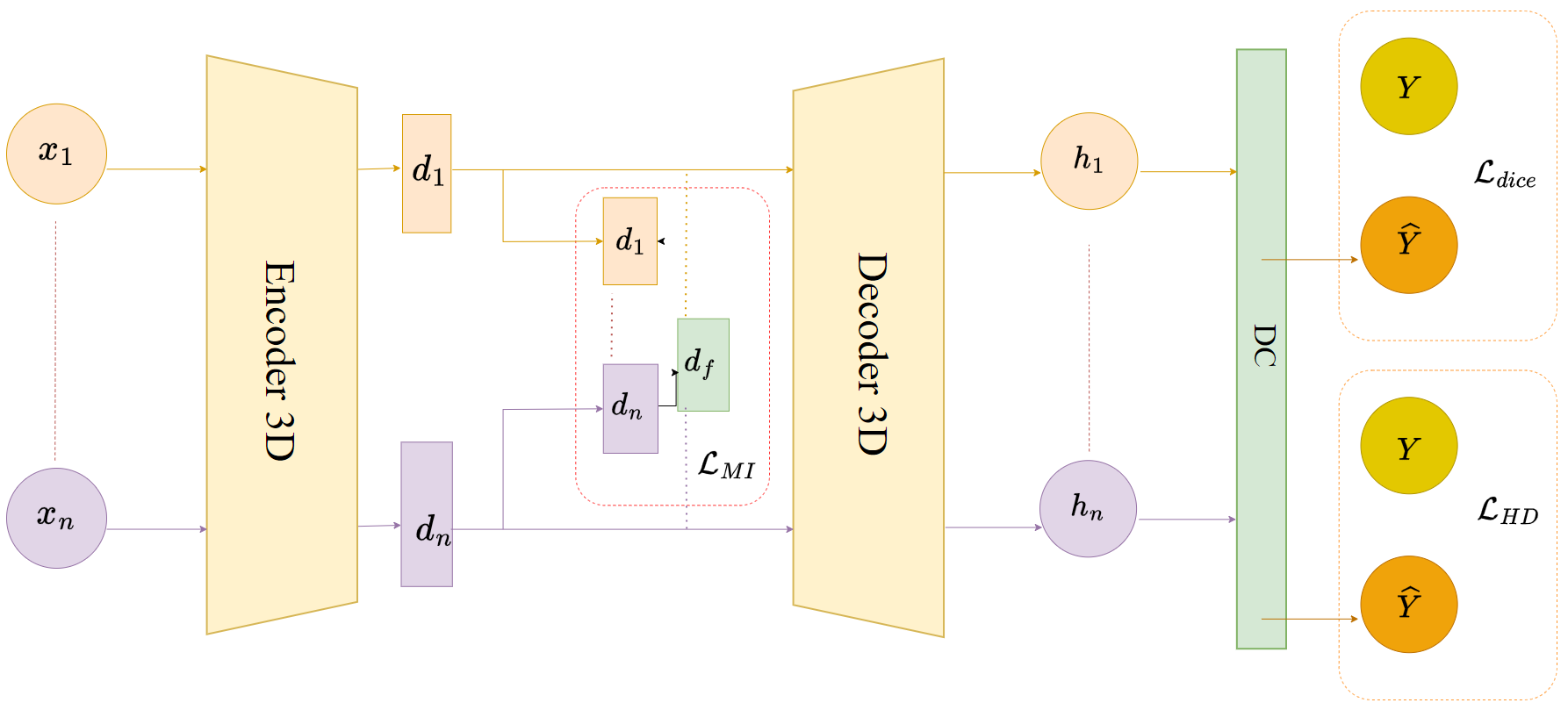}
	\caption{The Framework of Robust Divergence Learning for Missing-Modality Segmentation. This figure illustrates the overall structure of the proposed robust divergence learning approach, specifically designed to address segmentation challenges in scenarios where certain modalities are missing.}
	\label{fig:_1}
\end{figure*}

\section{Methodology}

\subsection{Knowledge Distillation for Segmentation Using Hölder Divergence}

Brain tumor segmentation, particularly glioma segmentation, involves distinguishing four categories: background, whole tumor, tumor core, and enhancing tumor. Missing modalities can degrade segmentation accuracy. The Hölder divergence is employed for its flexibility and robustness, making it suitable for complex models and non-symmetric data. It supports brain tumor segmentation under missing modalities, maintaining high accuracy in clinical settings.

The loss function using Hölder divergence is:
\begin{equation}
\frac{1}{D\times H\times W}\sum_{dhw} D_{\alpha}^{H}(\sigma(\mathbf{S}^{p}_{dhw})|\sigma(\mathbf{S}^{l}_{dhw})),
\end{equation}
where $\mathbf{S}^{p}_{dhw}$ and $\mathbf{S}^{l}_{dhw}$ are predicted and label probabilities for pixel $(d, h, w)$, and $D_{\alpha}^{H}$ denotes Hölder divergence, the definition is shown in Definition \ref{def_1}:
\newtheorem{definition}{\bf{Definition}}
\begin{definition} 
	\label{def_1}
	(\textbf{Hölder Statistical Pseudo-Divergence, HPD \cite{37}}) HPD pertains to the conjugate exponents $\alpha$ and $\beta$, where $\alpha \beta>0$. In the context of two densities, $p(x) \in {L^\alpha }\left( {\Omega,\nu } \right)$ and $q(x) \in {L^\beta }\left( {\Omega ,\nu } \right)$, both of which belong to positive measures absolutely continuous with respect to $\nu$, HPD is defined as the logarithmic ratio gap, as follows: $D_{\alpha}^{H}(p(x):q(x))=-\log\left(\frac{\int_{\Omega}p(x)q(x)\mathrm{d}x}{\left(\int_{\Omega}p(x)^{\alpha}\mathrm{d}x\right)^{\frac1\alpha}\left(\int_{\Omega}q(x)^{\beta}\mathrm{d}x\right)^{\frac1\beta}}\right),$ when $0<\alpha<1$ and $\beta  = \bar \alpha  = \frac{\alpha }{{\alpha  - 1}} < 0$ or $\alpha<0$ and $0<\beta<1.$
\end{definition}

\subsection{High Mutual Information Knowledge Transfer Learning}

In clinical practice, the challenge of missing modality segmentation often leads to incomplete information, limited model generalization, and data application constraints. To address these issues, a high mutual information knowledge transfer learning strategy between full and missing modalities is introduced. This strategy maximizes existing modality information, compensating for the loss caused by missing data, thereby enhancing model accuracy and stability with incomplete datasets.

Our approach involves extracting $K$ pairs of feature vectors $\left\{ \left( d_f^{(k)}, d_m^{(k)} \right) \right\}_{k=1}^K$ from the encoder layers of both the full and missing modality paths. By calculating the entropy $H(d_f)$ and conditional entropy $H(d_f \mid d_m)$ for each pair, we derive the mutual information $MI(d_f;d_m)=H(d_f)-H(d_f \mid d_m)$, which shows how the full modality path reduces uncertainty given the missing modality information. To estimate these mutual information values accurately, a variational information maximization method \cite{p54} is employed.

We approximate the conditional distribution $p(d_f\mid d_m)$ with the variational distribution $q(d_f\mid d_m) $ to optimize the layer-wise mutual information. The optimization process is defined by the following loss function $\mathcal{L}_{\mathcal{MI}} $:
\begin{equation}
\begin{aligned}
-\sum_{k=1}^{K} \gamma_{k} \mathbb{E}_{d_f^{(k)}, d_m^{(k)} \sim p\left( d_f^{(k)}, d_m^{(k)} \right)} 
                           \left[ \log q\left( d_f^{(k)} \mid d_m^{(k)} \right) \right],
\end{aligned}
\end{equation}

In our framework, the parameter $\gamma_{k} $ increases with the layer level $k$, reflecting the richer semantic information in higher network layers. This ensures effective knowledge transfer by assigning higher weights to these layers. The implementation of the variational distribution is given by:
\begin{equation}
        \begin{array}{l}
         - \log q(d_f\mid d_m)\begin{array}{*{20}{c}}
        {}&{}&{}&{}&{}&{}
        \end{array}\\
        \begin{array}{*{20}{c}}
        {}&{}
        \end{array} = \sum\limits_{c = 1}^C {\sum\limits_{h = 1}^H {\sum\limits_{w = 1}^W {\left( {\log {\sigma _c} + \frac{{{{\left( {{d_f^{c,h,w}} - {\mu^{c,h,w}}(d_m)} \right)}^2}}}{{2\sigma _c^2}}} \right)} } } \\
        \begin{array}{*{20}{c}}
        {}&{}
        \end{array} + {\rm{constant}},\begin{array}{*{20}{c}}
        {}&{}&{}&{}&{}&{}
        \end{array}
        \end{array}
\end{equation}
where \( \mu(\cdot) \) and \( \sigma \) represent the heteroscedastic mean and homoscedastic variance of the Gaussian distribution, respectively. \( W \) and \( H \) denote the width and height of the image, \( C \) represents the number of channels, and \( \mathrm{constant} \) is a fixed term.




\subsection{Overall Framework}
 Let ${X}$ and ${Y}$ denote samples from a multimodal dataset, where ${X}=\{x_j\}_{j=1}^M$ contains $M$ samples with $N$ modalities per sample: $x_j=\{x_j^{i}\}_{i=1}^N$, with $x_j^{i}\in {X}$ representing the $i^{th}$ modality data of the $j^{th}$ sample. Corresponding label is ${y}_j$. To leverage features from different modalities, a parallel 3D U-Net-based network is designed. Each sample's data is first encoded into a common feature space by channel encoders $ f_i$, unifying data representation across channels. Subsequently, each modality's data is independently input into a parameterized shared backbone $T(\cdot;\theta)$ to generate unique single-modality representations $h_i$: $ h_i=T(f_i(x_i);\theta).$

The final output in our shared network architecture includes a Dynamic Combination Network Module (DC). For missing modalities, we exclude their representations and use a flexible fusion operator \( {M}(\cdot) \) to integrate remaining single-modality representations $ H \subseteq \{h_1, h_2, \ldots, h_n\} $: $\widehat{Y} ={M}(H).$

The Dice loss function \cite{p49} is utilized to optimize consistency between the fused predicted image \( \widehat{{Y}} \) and target labels \( {Y} \), ensuring precise pixel-wise training:
\begin{equation}
   \mathcal{L}_{Dice}(\widehat{Y},Y)=1-\frac{2}{J}\sum_{j=1}^{J}\frac{\sum_{i=1}^{I}\widehat{Y}_{i,j}Y_{i,j}}{\sum_{i=1}^{I}\widehat{Y}_{i,j}^{2}+\sum_{i=1}^{I}Y_{i,j}^{2}},
\end{equation}
where $ I $ is the total number of voxels and $ J $ is the number of classes, $ \widehat{Y}_{i,j} $ is the predicted probability of voxel $ i $ belonging to class $ j $, and $ Y_{i,j} $ is the one-hot encoded label.

Facing the challenges of medical image processing with missing modalities, inspired by Hinton's knowledge distillation \cite{p53}, the high mutual information knowledge transfer loss \( \mathcal{L}_{\mathcal{MI}} \) is introduced to enhance model accuracy. After that, the total loss can be obtained:
\begin{equation}
    \mathcal{L}_{all}= \mathcal{L}_{Dice}(\widehat{Y},Y) + \mathcal{L}_{\mathcal{MI}} + \mathcal{L}_{HD}(\widehat{Y},Y).
\end{equation}
\begin{table*}
\centering
\caption{Quantitative Evaluation of Segmentation Results (DSC $\uparrow$) on BraTS 2018. This table presents the quantitative results of segmentation performance, measured by the Dice Similarity Coefficient (DSC), on the BraTS 2018 dataset. The results provide a comparative evaluation of the effectiveness of different segmentation methods, where higher DSC values indicate better segmentation accuracy.}
\resizebox{\textwidth}{!}
{
\setlength{\tabcolsep}{3pt}
\begin{tabular}{c|c|ccccccccccccccc|c}
\toprule[1.0pt] 
Task & Methods & Fl & T2 & T1c & T1 & T2,Fl & T1c,Fl & T1c,T2 & T1,Fl & T1,T2 & T1,T1c &$\sim\mathrm{T}1$ &$\sim\mathrm{T}1c$ &$\sim\mathrm{T}2$ &$\sim\mathrm{Fl.}$  &Full & Avg. \\
\midrule[0.5pt]
\multirow{6}{*}{WT} 
&RFNet	&85.8	&85.1	&73.6	&74.8	&89.3	&89.4	&85.6	&89.0	&85.4	&77.5	&90.4	&90.0	&89.9	&86.1	&90.6	&85.5 \\
&mmFormer	&86.1	&81.2	&72.2	&67.5	&87.6	&87.3	&83.0	&87.1	&82.2	&74.4	&88.1	&87.8	&87.3	&82.7	&89.6	&82.9 \\
&M3AE	&88.7	&84.8	&75.8	&74.4	&89.9	&89.7	&86.3	&89 	&86.7	&77.2	&90.2	&89.9	&88.9	&85.7	&90.1	&85.8 \\
&MTI	&88.4	&86.6	&77.8	&78.7	&90.3	&89.5	&88.2	&89.7	&88.1	&81.8	&90.6	&89.7	&90.4	&88.4	&90.6	&87.3 \\
&GGDM	&89.3	&87.0	&79.9	&75.9	&90.7	&90.6 	&88.6	&90.2	&88.2 	&81.1	&\textbf{91.3}	&\textbf{91.0} 	&90.5	&87.9	&91.0 	&87.6  \\
&OUR	&\textbf{89.8}	&\textbf{88.2}	&\textbf{80.5}	&\textbf{78.8}	&\textbf{90.8}	&\textbf{90.7}	&\textbf{89.3}	&\textbf{90.4}	&\textbf{88.7}	&\textbf{82.0}	&\textbf{91.3}	&\textbf{91.0}	&\textbf{90.9}	&\textbf{89.2}	&\textbf{91.3}	&\textbf{88.2}  \\
\midrule[0.5pt]

\multirow{6}{*}{TC} 
&RFNet	&62.6	&66.9	&80.3	&65.2	&71.8	&81.6	&82.4	&72.2	&71.1	&81.3	&82.6	&74.0	&82.3	&82.9	&82.9	&76.0 \\
&mmFormer	&61.2	&64.2	&75.4	&56.6	&69.8	&77.9	&78.6	&65.9	&69.4	&78.6	&79.6	&71.5	&79.8	&80.4	&85.8	&73.0 \\
&M3AE	&66.1	&69.4	&82.9	&66.4	&70.9	&84.4	&84.2	&70.8	&71.8	&83.4	&84.6	&72.7	&84.1	&84.4	&84.5	&77.4 \\
&MTI	&66.7	&68.8	&81.5	&65.6	&71.8	&84.8	&84.8	&72.0	&72.3	&83.5	&85.8	&74.1	&85.2	&85.8	&85.9	&77.9 \\
&GGDM	&\textbf{77.3}	&76.3	&85.3	&58.1	&78.5	&\textbf{87.0} 	&\textbf{87.6}	&76.3	&76.8	&85.6	&\textbf{87.1}	&78.3	&86.5	&86.2	&85.8	&80.8 \\
&OUR	&76.2	&\textbf{77.6}	&\textbf{86.5}	&\textbf{72.6}	&\textbf{79.3}	&86.6	&87.2	&\textbf{78.6}	&\textbf{79.1}	&\textbf{86.9}	&\textbf{87.1}	&\textbf{80.1}	&\textbf{87.1}	&\textbf{87.4}	&\textbf{87.3}	&\textbf{82.6} \\
\midrule[0.5pt]

\multirow{6}{*}{ET} 
&RFNet	&35.5	&43.0	&67.7	&32.3	&45.4	&72.5	&70.6	&38.5	&42.9	&68.5	&73.1	&46.0	&71.1	&70.9	&71.4	&56.6 \\
&mmFormer	&39.3	&43.1	&72.6	&32.5	&47.5	&75.1	&74.5	&43.0	&45.0	&74.0	&75.7	&47.7	&75.5	&74.8	&77.6	&59.9 \\
&M3AE	&35.6	&47.6	&73.7	&37.1	&45.4	&75.0	&75.3	&41.2	&48.7	&74.7	&73.8	&44.8	&74.0 	&75.4	&75.5	&59.9 \\
&MTI	&40.5	&41.4	&75.7	&44.5	&48.3	&76.8	&77.7	&44.4	&47.7	&77.1	&76.6	&50.0 	&77.4	&78.5	&80.4	&62.5 \\
&GGDM	&47.4	&\textbf{53.4}	&81.6	&34.7	&55.2	&82.0	&82.6	&51.1	&54.7	&82.0	&82.1	&56.0	&82.2	&82.8	&82.1	&67.6 \\
&OUR	&\textbf{48.6}	&52.9	&\textbf{82.4}	&\textbf{48.7}	&\textbf{55.8}	&\textbf{83.0}	&\textbf{83.2}	&\textbf{53.8}	&\textbf{56.9}	&\textbf{82.8}	&\textbf{83.7}	&\textbf{58.4}	&\textbf{83.2}	&\textbf{83.5}	&\textbf{84.1}	&\textbf{69.4} \\
\bottomrule[1.0pt]
\end{tabular}
}
\label{tab-2}
\end{table*}

\section{Experiments and Analysis}
\subsection{Datasets and Evaluation Metrics}	
To improve the model's logical coherence, result reliability, and algorithm robustness in brain tumor segmentation, this study utilizes the BraTS 2018 and BraTS 2020 datasets\cite{p38}. These datasets are widely recognized in the field of medical imaging for multi-classification and segmentation tasks. They are extensive collections of multi-modal MRI scans (T1, T1Gd, T2, and FLAIR) from patients with high-grade and low-grade gliomas. Expertly annotated, these datasets mark tumor subregions like the enhancing tumor, peritumoral edema, and necrotic core. They are essential for advancing and validating automated brain tumor segmentation algorithms. To evaluate the effectiveness of our method, the Dice Similarity Coefficient (DSC)\cite{p49}, $\begin{aligned}
\mathrm{Dice}(P,G)=\frac{2\times|P\cap G|}{|P|+|G|}
\end{aligned}$ is used, which is a common performance metric in medical image analysis. The DSC measures the overlap between the model's output ($P$) and the ground truth ($G$). A higher Dice coefficient indicates better predictive performance.


\subsection{Training Details}	
In this study, a PyTorch-based framework \cite{p43} (version 2.3.0) is utilized for training all models on a server equipped with dual NVIDIA RTX A6000 GPUs. The standard 3D U-Net architecture \cite{p41} is adopted, featuring a single encoder-decoder parallel processing structure that incorporates residual blocks and group normalization techniques. During training, a batch size of 8 is set, and the Adam optimizer \cite{p42} is employed to update model parameters, starting with a learning rate of 0.0008 and a weight decay of 0.00001. Training is conducted for 600 epochs to ensure comprehensive learning and performance optimization. Post-training, thorough testing of the model is conducted under all possible channel dropout configurations.

\subsubsection{Compare Experimental Models}	
The comparative experimental models employed in this study are RFNet \cite{p44}, MMFormer \cite{p45}, MA3E \cite{p46}, MTI \cite{p47}, and GGDM \cite{p48}. Each model makes unique contributions to the field of missing modality segmentation, as outlined below. The results for RFNet, MMFormer, MA3E, and MTI are sourced from their respective original research papers, all adhering to the same experimental configuration as RFNet. Additionally, the GGMD method is tested, according to authors' code.  These models are outlined below:

1. \textbf{RFNet (Ding et al., ICCV 2021) \cite{p44}}: RFNet is a region-aware fusion network designed for the segmentation of brain tumors in scenarios with incomplete multi-modal data. 2. \textbf{MMFormer (Zhang et al., MICCAI 2022) \cite{p45}}: MMFormer is a multimodal medical transformer developed to improve brain tumor segmentation in scenarios involving incomplete multimodal data. 3. \textbf{MA3E (Liu et al., AAAI 2023) \cite{p46}}: M3AE is a multimodal representation learning approach for brain tumor segmentation that effectively handles missing modalities. 4. \textbf{MTI (Ting and Liu, JBHI 2024) \cite{p47}}: MTI is a multimodal transformer designed to enhance brain tumor segmentation using incomplete MRI data. 5. \textbf{GGMD (Wang et al., AAAI 2024) \cite{p48}}: GGMD is a method designed to enhance robustness in brain tumor segmentation when handling missing modalities.

Tables \ref{tab-2}--\ref{tab-3} showcase the performance of our research method on the BraTS 2018 and BraTS 2020 datasets, benchmarked against five state-of-the-art brain tumor segmentation techniques, and the symbol $ \sim(\cdot) $ in Tables \ref{tab-2}--\ref{tab-3} denotes the amissing of a specific modality, with optimal performance results highlighted in black across different tumor types. The results highlight our method's superior performance across all three evaluated tumor regions—Whole Tumor (WT), Tumor Core (TC), and Enhancing Tumor (ET)—achieving the highest average Dice Similarity Coefficient (DSC). Specifically, Table \ref{tab-2} demonstrates improvements of 0.6\% in WT, 1.8\% in TC, and 1.8\% in ET regions compare to existing state-of-the-art methods on the BraTS 2018 dataset. Similarly, Table \ref{tab-3} shows enhancements of 0.8\% in WT, 0.7\% in TC, and 0.9\% in ET regions on the BraTS 2020 dataset.

\begin{figure}
	\centering
	\includegraphics[width=1.0\linewidth]{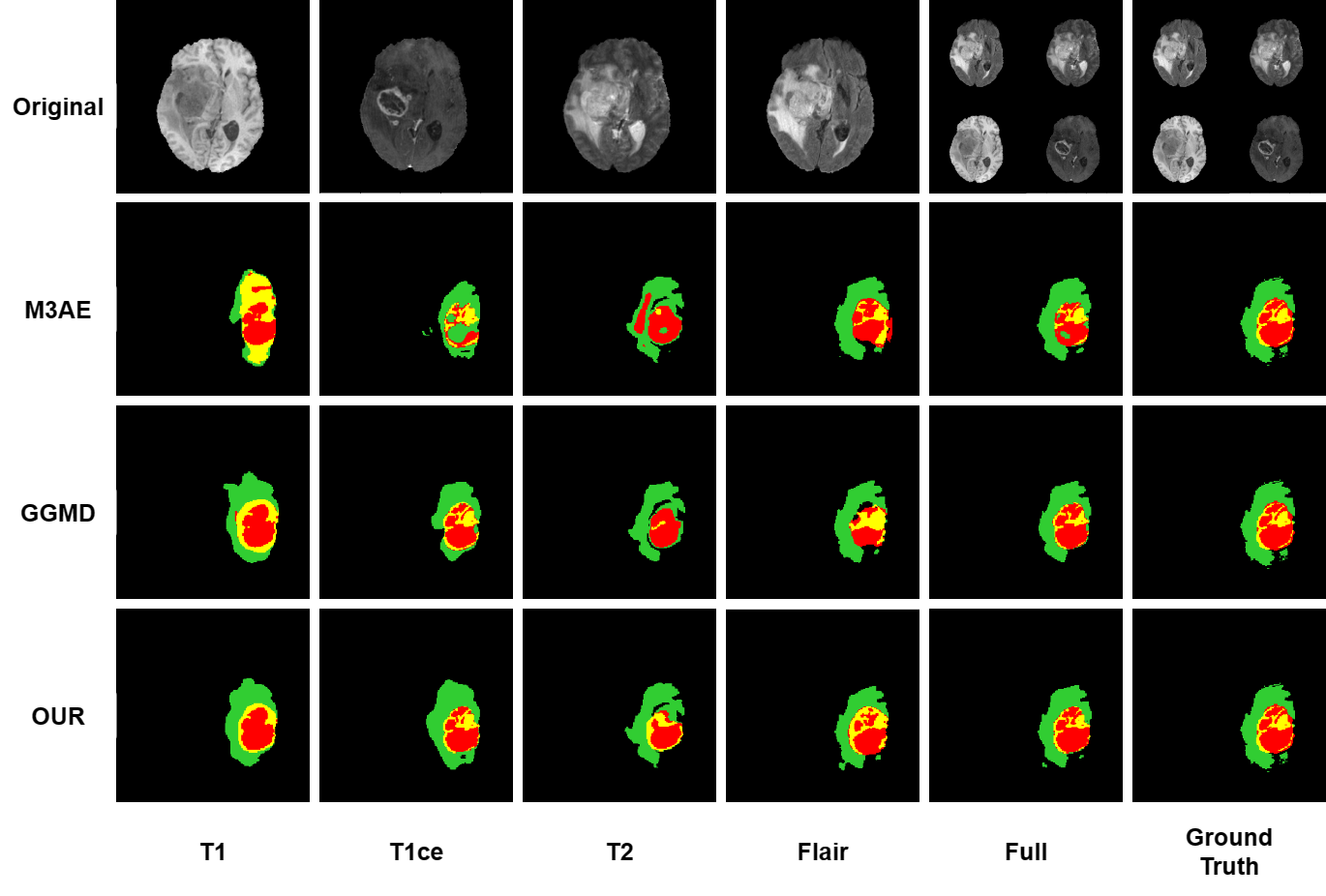}
	\caption{This figure presents the segmentation results of three models on the BraTS 2018 dataset using different modality inputs. The second row shows the reproduced results of the M3AE, the third row shows the reproduced results of the GGMD, and the fourth row displays the results of ours. Each column represents different input settings: the first four columns show the results for single modality inputs (T1, T1ce, T2, and Flair, respectively), the fifth column displays the results using all four modalities as input simultaneously, and the last column shows the corresponding ground truth. }
	\label{fig:_2}
\end{figure}

Our method excels notably in scenarios with missing multimodal data, achieving substantial gains of 2.0\% to 6.0\% in Dice coefficients. For instance, in Table \ref{tab-2}, when only the T1 modality is available, our method outperforms other advanced algorithms by 0.1\%, 6.2\%, and 4.2\% in WT, TC, and ET regions, respectively. Correspondingly, Table \ref{tab-3} indicates improvements of 2.4\%, 2.9\%, and 3.0\% under similar conditions.

These findings underscore the robustness of our method in maintaining efficient segmentation performance despite significant deficits in multimodal data.Furthermore, Figure \ref{fig:_2} compares our method with other advanced techniques like M3AE and GGMD under single-modal and full-modal input conditions. These segmentation results validate our conclusions from the evaluation metrics, demonstrating that our method surpasses other advanced technologies in handling multimodal data, especially when only a single modality is available. This performance advantage is particularly notable under single-modal input conditions.
 \begin{table*}
\centering
\caption{Quantitative Evaluation of Segmentation Results (DSC $\uparrow$) on BraTS 2020. This table provides a quantitative assessment of the segmentation performance on the BraTS 2020 dataset, measured using the Dice Similarity Coefficient (DSC). An upward arrow ($\uparrow$) indicates that higher DSC values correspond to better segmentation accuracy, allowing a clear comparison of the effectiveness of different models or approaches on this dataset.}
\resizebox{\textwidth}{!}
{
\setlength{\tabcolsep}{3pt}
\begin{tabular}{c|c|ccccccccccccccc|c}
\toprule[1.0pt] 
Task & Methods & Fl & T2 & T1c & T1 & T2,Fl & T1c,Fl & T1c,T2 & T1,Fl & T1,T2 & T1,T1c &$\sim\mathrm{T}1$ &$\sim\mathrm{T}1c$ &$\sim\mathrm{T}2$ &$\sim\mathrm{Fl.}$ & Full & Avg. \\
\midrule[0.5pt]
\multirow{6}{*}{WT} 
&RFNet	&87.3	&86.1	&76.8	&77.2	&89.9	&89.9	&87.7	&89.7	&87.7	&81.1	&90.7	&90.6	&90.7	&88.3	&91.1	&87.0 \\
&mmFormer	&86.5	&85.5	&78.0	&76.2	&89.4	&89.3	&87.5	&88.7	&86.9	&80.7	&90.4	&89.8	&89.7	&87.6	&90.5	&86.4 \\
&M3AE	&86.5	&86.1	&73.9	&76.7	&89.3	&89.5	&87.4	&89.4	&87.2	&78.1	&90.2	&90.4	&90.0	&88.6	&90.6	&86.3 \\
&MTI	&89.1	&86.5	&77.4	&78.1	&90.5	&90.0	&88.4	&89.9	&88.0	&81.2	&90.6	&90.3	&90.7	&88.7	&90.6	&87.3 \\
&GGDM  &91.0	&88.3	&80.6	&77.4	&92.1	&91.9	&89.8	&91.6	&89.3	&82.7	&92.3	&92.1	&91.6	&89.2	&92.0 	&88.8 \\
&OUR    &\textbf{91.9}	&\textbf{89.2}	&\textbf{81.5}	&\textbf{80.5}	&\textbf{92.5}	&\textbf{92.4}	&\textbf{90.0}	&\textbf{92.1}&\textbf{89.8}	&\textbf{83.2}	&\textbf{92.8}	&\textbf{92.6}	&\textbf{92.3}&\textbf{89.9}	&\textbf{92.7}	&\textbf{89.6}	\\
\midrule[0.5pt]

\multirow{6}{*}{TC} 
&RFNet	&69.2	&71	 &81.5	&66.0	&74.1	&84.7	&83.5	&73.1	&73.1	&83.4	&85	&75.2	&85.1	&83.5	&85.2	&78.2 \\
&mmFormer	&64.6	&63.3	&81.5	&63.2	&70.3	&83.7	&82.6	&71.7	&67.7	&82.8	&83.9	&72.4	&84.4	&79.0	&84.6	&75.7 \\
&M3AE	&68.0	&70.3	&81.4	&66.0	&75.0	&82.0	&83.0	&73.8	&72.5	&82.4	&83.1	&75.1	&82.4	&84.1	&84.4	&77.6 \\
&MTI	&69.3	&71.5	&83.4	&66.8	&75.5	&85.6	&86.4	&73.9	&73.3	&85.2	&86.4	&75.9	&86.5	&86.5	&87.4	&79.6 \\
&GGDM	&77.0	&79.1	&\textbf{87.6}	&70.5	&81.6	&88.0	&88.5	&80.3	&\textbf{81.1}	&88.0	&88.1	&82.4	&87.9	&88.4	&88.0 	&83.8 \\
&OUR	&\textbf{78.6}	&\textbf{79.8}	&87.5	&\textbf{73.4}	&\textbf{82.5}	&\textbf{88.9}	&\textbf{88.6}	&\textbf{81.2}	&80.8	&\textbf{88.1}	&\textbf{88.8}	&\textbf{82.5}	&\textbf{88.9}	&\textbf{88.6}	&\textbf{88.8}	&\textbf{84.5}  \\
\midrule[0.5pt]

\multirow{6}{*}{ET} 
&RFNet	&38.2	&46.3	&74.9	&37.3	&49.3	&76.7	&75.9	&41.0	&45.7	&78.0	&77.1	&49.9	&76.8	&77.0	&78.0	&61.5 \\
&mmFormer	&36.6	&49.0	&78.3	&37.6	&49.0	&79.4	&77.2	&42.9	&49.1	&81.7	&78.7	&50.0	&80.6	&68.3	&79.9	&62.6 \\
&M3AE	&40.5	&46.0	&72.4	&39.9	&47.3	&74.7	&76.8	&43.2	&46.6	&75.4	&77.1	&48.2	&75.9	&77.4	&78.0	&61.3 \\
&MTI	&43.6	&45.6	&78.9	&41.3	&48.7	&81.8	&81.7	&48.2	&50.0	&79.2	&81.0	&52.5	&81.8	&78.5	&81.6	&65.0 \\
&GGDM	&49.8	&52.5	&84.2	&39.7	&56.5	&84.6	&84.5	&54.6	&\textbf{55.3}	&84.2	&84.2	&\textbf{58.6}	&84.3	&\textbf{84.3}	&84.1	&69.4  \\
&OUR	&\textbf{51.9}	&\textbf{54.6}	&\textbf{84.6}	&\textbf{44.3}	&\textbf{57.7}	&\textbf{84.9}	&\textbf{84.8}	&\textbf{55.4}	&55.1	&\textbf{84.3}	&\textbf{84.9}	&\textbf{58.6}	&\textbf{84.4}	&\textbf{84.3}	&\textbf{84.3}	&\textbf{70.3}  \\
\bottomrule[1.0pt]
\end{tabular}
}
\label{tab-3}
\end{table*}

\begin{table}
\centering
\caption{Comparison of the Superiority of Hölder Divergence with Different $f$-Divergences on the Brats 2018 Dataset. This table illustrates the comparative performance of Hölder divergence against different types of $f$-divergences on the BraTS 2018 dataset.}
\setlength{\tabcolsep}{10pt}
\begin{tabular}{c|ccc|c}
\toprule[1.0pt]
\multicolumn{1}{c}{\textbf{Methods}} & \multicolumn{3}{c}{\textbf{Dice}} & \multicolumn{1}{c}{\textbf{}} \\
\midrule[0.5pt]
\textbf{$f$-divergence }     &\textbf{WT} & \textbf{TC} & \textbf{ET}  &\textbf{Avg.} \\
\midrule[0.5pt]
\textbf{Total Variation} \cite{p50}         &67.2  &1.9   &0.9   &23.3   \\
\textbf{Squared Hellinger } \cite{p50}       &85.3  &75.4  &60.1  &73.6  \\
\textbf{Kullback-Leibler} \cite{p50}               &84.5  &76.2  &61.4  &74.0  \\
\textbf{Neyman $\chi^2$} \cite{p50}        &83.4  &75.1  &59.9  &72.8  \\
\textbf{Jensen-Shannon } \cite{p50}         &84.6  &76.5  &59.8  &73.6  \\
\textbf{Hölder} \cite{37}                 &$\textbf{88.2}$  &$\textbf{82.6}$  &$\textbf{69.4}$  &$\textbf{80.1}$  \\
\bottomrule[1.0pt]
\end{tabular}
\label{tab-4}
\end{table}

\subsubsection{Exploration of the Superiority of Hölder Divergence}	
To explore the superiority of Hölder divergence, this study conduct experimental comparisons using Hölder divergence and other $f$-divergences \cite{p50}, including Total Variation \cite{p50}, Squared Hellinger \cite{p50}, Kullback-Leibler \cite{p50}, Neyman $\chi^2$ \cite{p50}, and Jensen-Shannon divergence \cite{p50}. As shown in the table \ref{tab-4}, the average Dice coefficient of Hölder divergence reached 80.1\% after adjusting the hyperparameter $\alpha$, which is 6.1\% higher than the best-performing alternative methods. This significant performance advantage underscores the importance of Hölder divergence in improving model accuracy.

The experimental data consistently demonstrate that the application of Hölder divergence significantly enhances segmentation task performance, validating its effectiveness in the field of medical image processing. Additionally, our research reveals the critical role of Hölder divergence in enhancing knowledge distillation techniques to improve segmentation efficiency, providing valuable references and guidance for future research and development in related technologies. These findings not only deepen our understanding of the potential of Hölder divergence but also provide empirical evidence for optimizing deep learning models using this method.
\begin{table}
\centering
\caption{Exploring the Impact of Hölder Conjugate Exponents on Experimental Results Based on the Brats 2018 Dataset. This table illustrates how varying the Hölder conjugate exponents affects the experimental outcomes derived from the BraTS 2018 dataset. It highlights the relationships between different conjugate exponent values and their influence on the performance metrics within the context of the experiments.}
\setlength{\tabcolsep}{10pt}
\begin{tabular}{cc|ccc|c}
\toprule[1.0pt]
\multicolumn{2}{c}{\textbf{Methods}} & \multicolumn{3}{c}{\textbf{Dice}} & \multicolumn{1}{c}{\textbf{}} \\
\midrule[0.5pt]
\textbf{divergence }  &$\bm{\alpha}$    &\textbf{WT} & \textbf{TC} & \textbf{ET}  &\textbf{Avg.} \\
\midrule[0.5pt]
-                &-                &87.8  &82.9  &67.4  &79.4    \\
\textbf{KL}      &-                &84.5  &76.2  &61.4  &74.0  \\
\textbf{Hölder}  &1.05    &87.8  &82.6  &69.2  &79.9  \\
\textbf{Hölder}  &1.08    &88.2  &82.6  &68.7  &79.9  \\
\textbf{Hölder}  &\textbf{1.10}   &\textbf{88.2} &82.6  &\textbf{69.4}  &\textbf{80.1}  \\
\textbf{Hölder}  &1.15   &88.1  &82.8  &67.9  &79.6   \\
\textbf{Hölder}  &1.20   &87.9  &\textbf{83.1} &68.1  &79.7  \\
\bottomrule[1.0pt]
\end{tabular}
\label{tab-5}
\end{table}

\subsection{Exploring the Impact of Hölder Conjugate Exponents on Experimental Results}
To further investigate the impact of Hölder conjugate exponents on experimental results, we explore various Hölder conjugate exponents. As shown in Table \ref{tab-5}, this study compares the performance under different Hölder hyperparameters ($\alpha$), KLD, and without the application of knowledge distillation. The experimental results indicate that when the Hölder divergence hyperparameter $\alpha = 1.1$, the performance improves by an average of 0.7\% compared to the case without knowledge distillation and by 6.1\% compared to KL divergence. This result underscores the crucial role of selecting an appropriate Hölder conjugate exponent ($\alpha$) in significantly enhancing model performance. 

Our findings clearly demonstrate the critical role of Hölder divergence in enhancing knowledge distillation techniques to improve segmentation task efficiency. Throughout our experiments, we consistently observe that setting the Hölder conjugate exponent to $\alpha = 1.1$ markedly improves the model's segmentation performance, further validating the effectiveness of Hölder divergence.
\begin{table}
\centering
\caption{Quantitative evaluation results of the ablation study on the BraTS 2018 dataset. This table presents the impact of different model components on performance, highlighting the effectiveness of each component in contributing to overall accuracy and robustness.}
\setlength{\tabcolsep}{6pt}
\begin{tabular}{ccc|cccc|c}
\toprule[1.0pt]
\multicolumn{3}{c}{\textbf{Methods}} & \multicolumn{4}{c}{\textbf{Number of Missing Modalities}} & \multicolumn{1}{c}{\textbf{}} \\
\midrule[0.5pt]
$\mathcal{L}_{dice}$  &$\mathcal{L}_{MI}$  &$\mathcal{L}_{HD}$  &\textbf{3} & \textbf{2} & \textbf{1} & \textbf{0} & \textbf{Avg.} \\
\midrule[0.5pt]
$\checkmark$   &$\times$  &$\times$  &60.2  &71.9  &77.1  &80.5  &70.7  \\
$\checkmark$  &$\checkmark$   &$\times$ &72.4  &79.7  &83.9  &87.0  &79.4  \\
$\checkmark$   &$\times$ &$\checkmark$  &72.9  &80.1  &84.3  &87.4  &79.8 \\
$\checkmark$  &$\checkmark$   &$\checkmark$   & \textbf{73.6}  & \textbf{80.3} & \textbf{84.4}  & \textbf{87.6} & \textbf{80.1} \\
\bottomrule[1.0pt]
\end{tabular}
\label{tab-6}
\end{table}
\subsection{Ablation Study}

In this study, we conduct a series of ablation experiments to demonstrate the effectiveness of mutual information knowledge transfer, denoted as $\mathcal{L}_{MI}$, between full and missing modalities. Additionally, we explore the impact of the Hölder divergence-based loss function, $\mathcal{L}_{HD}$, on model performance. Compared to traditional multimodal processing methods, our parallel network framework selectively activates only the data relevant to the available modalities. This approach effectively preserves the unique information of each modality, enhancing the model's ability to recognize diverse data features. Furthermore, the introduction of the Hölder divergence loss function improves the model's performance in handling multimodal data by precisely quantifying the mutual information between modalities, thereby promoting better feature alignment.

First, we evaluate the utility of different components within our network architecture, including the Dice loss function $\mathcal{L}_{Dice}$, mutual information knowledge transfer $\mathcal{L}_{MI}$ between full and missing modalities, and Hölder divergence-based knowledge distillation $\mathcal{L}_{HD}$. The results, as shown in Table \ref{tab-6}, indicate that the mutual information knowledge transfer $\mathcal{L}_{MI}$ and Hölder divergence-based knowledge distillation $\mathcal{L}_{HD}$ effectively improve model performance in various missing modality scenarios compared to the traditional segmentation loss $\mathcal{L}_{dice}$ alone. Specifically, when three modalities are missing, these strategies improve performance by 12.2\% and 12.7\%, respectively; with two missing modalities, they improve performance by 7.8\% and 8.2\%; with one missing modality, they improve performance by 6.8\% and 7.2\%; and with all modalities present, they improve performance by 6.5\% and 6.9\%. On average, the performance improvement for different modality inputs are 8.7\% and 9.1\%. Moreover, the combination of the parallel network architecture, mutual information knowledge transfer between full and missing modalities, and Hölder divergence-based knowledge distillation achieve the best results, further validating the effectiveness and superiority of our approach.

\subsection{Conclusion}
In this work, we present the quantitative evaluation results of our proposed method for addressing missing modality segmentation, a common challenge in clinical practice. Our approach utilizes a 3D U-Net combined with a parallel network architecture, integrating mutual information knowledge transfer and knowledge distillation based on Hölder divergence. This method enhances brain tumor segmentation capabilities despite missing modalities by efficiently transferring knowledge and optimizing model generalization. Key components of our framework include: 1. Parallel U-Net network with single modality input to handle missing modalities. 2. Mutual information knowledge transfer to enhance model processing capabilities. 3. Optimization of knowledge transfer efficiency through Hölder divergence during knowledge distillation. Ablation experiments highlight the critical role of each component and reveal the impact of the Hölder conjugate exponent on model performance.

Despite its effectiveness, the method has some limitations: 1. Training costs due to the large number of parameters and extensive tuning requirements, and 2. Sensitivity to hyperparameter selection, necessitating extensive experimentation and validation.
\section*{Acknowledgement}
This research was supported by Guangdong Basic and Applied Basic Research Foundation (No.2024A1515011774), the National Key Research and Development Program of China (No. 2022YFC3310300), and Beijing Natural Science Foundation (No. Z210001).

\ifCLASSOPTIONcaptionsoff
\newpage
\fi
	
\bibliographystyle{IEEEtran}
\bibliography{IEEEabrv,references.bib}

\end{document}